# ON THE OPTIMIZATION OF BITTORRENT-LIKE PROTOCOLS FOR INTERACTIVE ON-DEMAND STREAMING SYSTEMS

Carlo Kleber da Silva Rodrigues[1,2]

[1]Electrical and Electronics Department, Armed Forces University – ESPE, Sangolquí, Ecuador
[2]University Center of Brasilia, Brasília, DF, Brazil

## ABSTRACT

*This paper proposes two novel optimized BitTorrent-like protocols for interactive multimedia streaming: the Simple Interactive Streaming Protocol (SISP) and the Exclusive Interactive Streaming Protocol (EISP). The former chiefly seeks a trade-off between playback continuity and data diversity, while the latter is mostly focused on playback continuity. To assure a thorough and up-to-date approach, related work is carefully examined and important open issues, concerning the design of BitTorrent-like algorithms, are analyzed as well. Through simulations, in a variety of near-real file replication scenarios, the novel protocols are evaluated using distinct performance metrics. Among the major findings, the final results show that the two novel proposals are efficient and, besides, focusing on playback continuity ends up being the best design concept to achieve high quality of service. Lastly, avenues for further research are included at the end of this paper as well.*

## 1. INTRODUCTION

The BitTorrent protocol [1] is acknowledged as one of the most successful peer-to-peer (P2P) solutions for replicating content on the Internet [2]. Leveraging the upload bandwidth of the peers is the key principle. More precisely, the wanted content file is divided into data pieces, and the piece-replication work is then shared among the peers. The more data pieces a peer uploads to other peers, the more data pieces it may receive from these peers. Moreover, there is no dependence on multicast service, content distribution networks (CDN), main central data servers, or proxies [3]. Compared to traditional solutions, the system infrastructure complexity is reduced and the traffic load is more evenly distributed among the network nodes [4].

Being aware of the BitTorrent's success, researchers from both academy and industry have proposed to adapt it to on-demand multimedia streaming on the Internet. To this end, the piece-selection policy and the peer-selection policy, which form the core of this protocol, are modified to guarantee a satisfactory quality of service (QoS) to the end client who downloads the content [2, 5–9]. The piece-selection policy determines which data pieces should be requested by a peer, while the peer-selection policy dictates which peers should be chosen to receive the data pieces a peer owns.

Despite the noticeable efforts already done towards the above adaptation, there are at least four important open issues that have been the focus of recent research: (i) the peer request problem [10]; (ii) the service scheduling problem [10]; (iii) the client's heterogeneity [11–13]; and (iv) the number of concurrent data upload slots [14, 15]. Achieving solutions to these issues allows a more efficient design of novel BitTorrent-like protocols.

The above context gives the motivation for this paper, whose major goal is to propose two optimized BitTorrent-like protocols targeted at multimedia streaming: the *Simple Interactive Streaming Protocol* (SISP) and the *Exclusive Interactive Streaming Protocol* (EISP). The former chiefly seeks a trade-off between playback continuity and data diversity, while the latter is mostly focused on playback continuity. Two general steps are followed to assure a thorough and up-to-date approach. First, related work is carefully examined to identify positive features that may be included in the design of the novel proposals. Second, the four issues mentioned above are analysed to obtain solutions to be implemented on the novel proposals.

Performance evaluation is conducted through simulations on distinct file replication scenarios. Three client's interactivity profiles and four popular content types are considered, enabling a wide-spectrum analysis. *Retrieval Coefficient* (RC), *Interruption Coefficient* (IC), *Relativized Service Time* (RST), and *Number of Served Clients* (CS) are the metrics defined and assessed in the simulations. They are able reflect the QoS experienced by the system client. Among the main findings, the results show that the novel proposals are efficient and that prioritizing playback continuity is the most adequate design concept to achieve high system QoS.

The remainder of this text is structured as follows. Section 2 brings the basis to understand the subsequent sections. In Section 3, related work is examined. The open issues of BitTorrent-like algorithms are explored in Section 4. Section 5 presents the two novel proposals. Performance evaluation constitutes Section 6. Lastly, Section 7 summarizes some of the prime conclusions and suggests avenues for further research as well.

## 2. BASIS

### 2.1. BitTorrent overview

Understanding BitTorrent demands being aware of the key components involved in the file replication process: the tracker, the swarm, the neighbourhood set, the peer-selection policy, the piece-selection policy, the leechers, and the seeds [1, 16, 6]. These components and their corresponding roles are succinctly explained in what follows.

To receive a file, a newcomer (i.e., a new peer) first contacts the tracker, which is the system central entity that coordinates the communication among the peers already participating in the replication. These peers constitute a group denoted as *swarm*. After being contacted, the tracker provides the newcomer with a list $L$ containing the swarm peers. The newcomer then tries to establish bidirectional TCP connections with these peers. Those connections resulting successful form the neighbourhood set of the newcomer. This set contains the peers with which the newcomer may have data sharing, i.e., from which it may download data pieces and to which it may upload data pieces.

There are two types of peers: *leecher* and *seed*. The former is a peer that is still downloading content, but also lets other peers download content from it. The latter is a peer that already has all data pieces, i.e., the whole content file, but stays in the system just to let other peers download content from it. The exchange of data pieces among peers is governed by the peer and piece selection policies.

The peer-selection policy, denoted as the *choke algorithm* or the *tit-for-tat algorithm*, allows a swarm peer $P$ to decide which peers of its neighbour set may receive the data pieces it owns. The peers that provide data to the peer $P$ at the highest data rates are selected. Typically, three peers are selected, and one data upload slot is assigned to each of them. These peers are said to be *unchoked*, and the other remaining ones are said to be *choked*. This evaluation is repeated at regular intervals, typically of 10 seconds, denoted as *regular unchoking intervals* [1, 16, 6].

There are the *optimistic unchoking intervals* as well. In this case, a swarm peer periodically, typically at every 30 seconds, selects at random another peer of its neighbour set to receive the data pieces it owns. One data upload slot is assigned to this selected peer. The optimistic unchoke has two purposes: (i) to evaluate the upload capacity of new peers, and (ii) to provide the new peers with their first piece as early as possible. The peer-selection policy is mainly designed to give upload incentives to the swarm peers and yield resilience to *free riders*, i.e., those peers willing to download pieces only [1, 16, 6].

By its turn, the piece-selection policy determines which data pieces a swarm peer should request next. For each swarm peer, a *rarest-pieces* set is defined. The *rarest pieces* of a peer are the less replicated pieces within its neighbourhood set. Right after being unchoked by a neighbour, the peer requests the next piece to download considering its rarest-pieces set and the available pieces on the neighbour that unchokes it. In doing so, the rarest piece requested by the peer may not be the globally rarest one. After receiving a piece, the peer then sends out a *piece-have* message to all its neighbours, informing about this piece. This policy is mainly designed to yield a fast piece replication and minimize file download times [1, 16, 6].

Lastly, as already mentioned, content files distributed by the BitTorrent protocol are first split into pieces. However, each piece is still split into blocks. Pieces are typically 256 kB in size, while blocks are typically 16 kB in size. Blocks are the data unit on the physical network, but the replication process only logically takes into account transferred data pieces. In particular, only complete pieces can be served by a peer [1, 16, 6].

## 2.2. Streaming Services and Interactivity

Under the P2P paradigm, the streaming services may be broadly classified into three types: file replication, on-demand, and live streaming [8, 10, 17, 18]. They basically differ in generation, distribution, and synchronization among the peers, as briefly explained in what follows.

In file sharing, the file has to be first generated and, only then, may be distributed. In this case, the file is available for playing only after the complete download by the client. In on-demand, the file also has to be first generated and, only then, distributed. Nevertheless, the file may be played by the client since the start of the download. Lastly, in live-streaming, clients are synchronized playing the file at the same instant and around a same file playback point. In this case, the file is generated and distributed simultaneously. It is worth noting that on-demand streaming possesses greater data diversity than live-streaming. This is because, in on-demand streaming, clients may request data at different times and hence their playback points are likely to differ greatly, whereas in live-streaming, clients are synchronized.

The BitTorrent protocol has already been proven cost-effective for file sharing and on-demand streaming. It is true though that some adaptation is necessary in the case of the latter due to the strict time constraints for the data being retrieved. More precisely, each file data piece has a deadline to be played. Under the BitTorrent paradigm, the data pieces are not requested in-order and hence the interarrival time between in-order consecutive pieces may greatly vary, resulting in a non-continuous file playback. Besides, it is very likely that it takes some waiting time until a newcomer may finally start receiving its first data pieces and may then begin the file playback. Hence, modifications in BitTorrent's core policies become crucial.

On the other hand, live streaming is still an intricate scenario for the deployment of BitTorrent. The data pieces do not exist a priori, since they are dynamically generated. Hence, adapting BitTorrent to this scenario requires, e.g., modifying the tracker entity to some extent as well as the communication process among peers. Thinking up new strategies for advertising, updating, and requesting new pieces turns out to be mandatory. As a matter of fact, implementing all these modifications would likely result in a quite new protocol, bearing very little resemblance to the traditional protocol.

Lastly, providing the client with interactivity is an intriguing challenge that has been faced by the latest on-demand streaming applications. In this case, the client may execute interactive actions (e.g., *Pause* and *Jump Forwards*) during the file playback, as though he were using a traditional DVD player. See that, during a non-interactive playback, the order the pieces are going to be requested and played by the client is known in advance: the playback starts at the first data piece and continues uninterruptedly up to the last data piece. In this case, there is a strict in-order playback. On the other hand, during an interactive playback, the playing order is not known in advance, since the client may freely choose the file parts he feels like playing at any time instant [3, 6, 9, 19]. This work is focused on interactive on-demand streaming.

## 3. RELATED WORK

This section is divided into two subsections. Subsection 3.1 debates some important proposals based on the modification of the BitTorrent's piece-selection policy. Subsection 3.2 focuses on proposals based on the modification of the BitTorrent's peer-selection policy. In case a same work modifies both piece and peer selection policies, it is then referenced in both subsections.

### 3.1. Piece-Selection Policy

The proposals modifying the piece-selection policy seek to obtain a trade-off between satisfying the time requirements of on-demand streaming and maintaining high piece diversity. They broadly classify in the following categories [20, 2]: *windows based*, *probability based*, *priority based*, and *optimization based*.

*Sliding window* is the key concept of the *windows-based* category. The sliding window may be seen as a dynamic piece set which indicates the file pieces to be requested with higher priority, usually because these pieces are near the client's current playback point and hence close in time to being played. As the file download progresses, the sliding window *slides* through the file to follow the latest downloaded data piece and encompass the nearest data pieces, thereby being dynamically updated. The window size and the number of windows may vary from proposal to proposal. Lastly, rarest-first or greedy (i.e., strict in-order) strategies are usually employed for requesting data pieces within it.

For example, Hoffmann et al. [3] define two fixed-sized sliding windows. One window contains the high-priority pieces, and the other contains pieces which are predicted by a client's behaviour model. The piece selection within both windows follows a rarest-first strategy, and the windows are chosen alternately by the requesting peer. The clients may execute interactive actions. Its main drawback is possibly the additional operation overhead resulting from the use of a client's behaviour model. Shah and Pâris [21] define a single fixed-sized window that contains high-priority pieces. These pieces are requested in accordance to a rarest-first policy. There is no support for interactivity, and operation simplicity may be highlighted as its main positive aspect. Lastly, Borghol et al. [20] define an adaptive window whose size changes in function of the amount of in-order data available in the peer's buffer and the peer's current playback position. When the window is small, in-order piece retrieval occurs, whereas when the window is large, rarest-first piece retrieval is considered. No interactivity is supported. The adaptive window and the flexible piece-selection policy are together the main positive aspect.

The *probability-based* category has the pieces selected according to a probability distribution. For example, Carlsson and Eager [22] propose an elaborated scheme in which each piece is assigned an exclusive request probability, computed from a Zipf distribution. No support for interactivity is provided. This category chiefly enables a significant freedom for deciding about the degree of specificity to be employed for selecting each file data piece.

The *priority-based* category divides the file into piece sets to more efficiently decide on the priority to request the pieces and on the retrieval policy to be used. The size of each set and the number of sets may vary from proposal to proposal. For example, Mol et al. [12] define three piece sets: high priority, mid priority, and low priority. The index of the requested piece and the requesting peer's current playback position determine the set the piece belongs to. Pieces of the high-priority set are requested before those of the mid-priority set, and pieces of the mid-priority set are requested before those of the low-priority set. A greedy selection is used within the high-priority set if the requesting peer has already started playback, rarest-first otherwise; and rarest-first selection is used within both mid-priority and low-priority sets. No support for interactivity is provided. Lastly, Vlavianos et al. [17] advocate that pieces belong to one of two piece sets: high-priority and low-priority. As expected, the former contains those pieces close in time to being played, and the latter refers to those which are needed in the future. These sets are chosen in function of predefined probabilities which may be adjusted to adapt to different requirements such as prioritizing playback continuity. The rarest-first policy is used for selecting pieces within both piece sets. No support for interactivity is included.

The *optimization-based* category thinks of the piece selection as an optimization problem. The goal is to discover an optimal order, among all possible ones, in which pieces may be downloaded. For example, Romero et al. [8] reduce the piece-selection policy to a combinatorial optimization, translated into a formulation of an *Asymmetric Traveling Salesman Problem* (ATSP), which is solved by means of a heuristic considering the *Ant Colony Optimization* (ACO). Time is discretized and an optimal order in which pieces may be downloaded is then determined. No support for interactive actions is provided. Even though a near-optimal solution may be reached, some computing overhead is introduced.

From above, the following points constitute the basis for designing efficient piece-selection policies: (i) the concept of sliding windows is the simplest solution for following non-sequential playback points which result from the execution of interactive actions; (ii) a greedy piece-selection policy is the most adequate for high-priority pieces, since it is the most common solution to achieve playback continuity; (iii) for selecting pieces other than the high-priority ones, a rarest-first policy should be employed, since it has been noted as the most effective strategy for providing piece diversity; (iv) dynamic adaptations (i.e., on-line computations) should be, whenever possible, avoided to prevent additional overheads from being introduced, what would be disastrous to system QoS.

## 3.2. Peer-selection policy

Designing peer-selection policies for P2P on-demand streaming systems must consider two relevant points. First, if the playback follows a strict sequential order (i.e., the playback starts at the first file data piece and continues uninterruptedly up to the last data piece), peers arriving later will always have a playback position behind, and hence no data that interest those peers arriving earlier. The incentives, provided by the original peer-selection policy, are not therefore the same as in traditional transfer. Second, having a download playback rate above the file playback rate does not bring any advantage at all. If this fact is ignored, as it occurs in the original peer-selection policy, there may be a situation in which there is enough aggregate upload capacity to serve all system peers, but not all peers have an acceptable QoS, since high-capacity peers are often favoured.

Thus, the prime challenge is to deal with the two above points while still providing incentives to swarm peers and resilience to free-riders. The most common approaches to face this broadly lie into four types of solutions: (i) use optimistic data upload slots more frequently; (ii) use indirect reciprocity; (iii) adjust the number of data upload slots a peer may open; (iv) consider the resulting data dispersion before selecting a peer. Within this context, several important proposals are discussed in the following.

The proposal of Shah and Pâris [21] is related to solution (i). It is defined a sliding window containing the next pieces to be consumed by the peer. At the beginning of every window playback, the peer performs a new optimistic unchoke. This results in more free tries to a larger number of system peers to download pieces which they can use to share later. So, it is very liely that newcomers may start participating in the file replication much earlier. Nevertheless, increasing the number of unchokes does not provide incentives for cooperation; besides, a peer may experience QoS degradation because of too much altruism.

The proposal of Mol et al. [12] is based on the principle of indirect reciprocity, and therefore fits in with solution (ii). For the three regular unchoke slots, each peer uses indirect reciprocity. Suppose a peer $P$ is up to unchoke three of its neighbours. All its neighbours are then sorted in function of the decreasing number of pieces they have forwarded to other peers than peer $P$, but counting only data which they originally received from peer $P$. The three first neighbours are then selected (i.e., unchoked) by peer $P$. To break ties, it is considered the total data forwarded to other peers than peer $P$. As for the optimistic unchoke slots, the time interval is typically set to 20 seconds. Exceptionally, if there is enough spare capacity, a peer may open two extra regular unchoke slots. The key drawback is the dependence on indirect information concerning data contributions. This makes the system more vulnerable to malicious peers than tit-for-tat schemes, where only direct information among peers is considered.

The work of D'Acunto et al. [11] belongs to solution (iii). Three schemes are introduced: the first employs mathematical formulas which consider, e.g., the peer's upload capacity and the playback rate; the second lets peers dynamically adjust the number of their optimistic slots in function of their current QoS; finally, the third scheme gives higher priority to newcomers during optimistic unchokes. The results show that this solution seems to be especially suitable for heterogeneous environments, where swarm peers own distinct upload and/or download capacities. On one hand, the proposed schemes cover an important range of strategies for deciding on the ideal number of data upload slots. On the other hand, they may unfortunately present an unacceptable operation overhead due to on-line computations.

Lastly, the work of Rocha and Rodrigues [19] is an example of solution (iv). It is proposed an algorithm for peer selection which has three questions as main guidelines: (1) How much do piece requests diverge from each other considering their arrival times? (2) How many of the retrieved file pieces may be effectively shared? (3) How often is each file data piece requested? Answering these questions makes it possible to compute the data-dispersion level a given neighbour peer introduces if unchoked. The less dispersion a neighbour introduces, the more efficiently the data retrieved from it may be shared. So, the algorithm is designed to guarantee the choice of neighbours which may minimize data dispersion. This solution is very likely to be the only one that considers the influences resulting from the retrieved data. Its chief drawback is the operation overhead due to the on-line computation.

From above, the following understandings constitute the basis for the design of piece-selection policies: (i) using more optimistic slots than regular slots becomes prohibitive, since altruism may deteriorate the system QoS; (ii) deploying the indirect-reciprocity principle may let the system vulnerable to malicious peers; (iii) introducing dynamic adaptations (i.e., on-line computations) causes additional overheads, what may be disastrous to system QoS.

## 4. ISSUES ON BITTORRENT-LIKE ALGORITHMS

In this section, four important open issues concerning BitTorrent-like algorithms for on-demand streaming are succinctly reviewed. The goal is to understand their essentialities and then devise solutions to be implemented on the novel proposals.

### 4.1. Peer Request Problem and Service Scheduling Problem

The *peer request problem* [10] may be understood from the following observation. A swarm peer may send requests for the rarest data pieces to the neighbours unchoking it. It is though very likely that these neighbours have common pieces. Nevertheless, under *normal operation* (please see Section 6.2), a same piece is not requested to more than one unchoking neighbour. So, it is necessary to choose one unchoking neighbour, among all those having the same piece.

Among the solutions introduced by Yang et al. [10], the *Least Loaded Peer Piggyback* (LLP-P) deserves more attention due to its attractive performance. Each peer chooses the unchoking neighbour with the shortest queue size, randomly breaking ties. For this implementation, each peer reports its queue length to all of its neighbours by explicit periodic messages and piggybacking on *piece-have* messages. In doing so, requests are assured to be evenly distributed among peers, resulting in a well-balanced request load and improving the system playback quality. Similar results are observed under mixed piece selections (e.g., greedy and rarest-first) and under heterogeneous environments. Its main drawback is the reasonable message overhead.

Now, see that the number of data-piece requests a peer may concurrently receive is a function of the number of neighbours it has. Nonetheless, there is a limit for serving these requests concurrently. This limit is determined by the number of upload slots a peer has. Each peer is usually configured to own four slots. Hence, in case more than four neighbours are allowed to make requests, there will be requests waiting in queues until a slot becomes available. The playback deadlines of the requested data pieces are quite diverse, some having urgent deadlines and others being more relaxed. So, the question is which request in its queue a peer should serve next. Reducing the probability of a data piece missing its playback deadline is hence the chief goal. This is formally denoted as the *service scheduling problem* [10]. Another related question is whether all queued requests for data pieces should be served; in case the answer is negative, which ones should then be served.

Yang et al. [10] use deadline-aware scheduling to solve the above. Two policies are used in conjunction: *Earliest Deadline First* (EDF) and *Early Drop* (ED). Under EDF, a peer maintains its queue sorted by the request deadline, and picks the request with the most urgent deadline to serve, randomly breaking ties. Under ED, a peer estimates the waiting time of a newly arrived request, using the currently available bandwidth and the request load already in the queue. If it is concluded that the newly arrived request can make its deadline, it is inserted into the queue; otherwise, it is simply dropped. Inserting a newly arrived request makes it necessary to estimate the waiting time of all requests that were already in the queue but ended up being placed behind the newly arrived request. Requests then missing their deadlines are dropped from the queue.

Even though the above solution is certain to generate significant message overhead, since pieces requests have to be reissued when requests are dropped by a peer, the results show that the overall quality may still be improved a little. Similar observations apply under mixed piece selections as well. On the other hand, and maybe worst of all, to achieve a real satisfactory system QoS, it is mandatory to simultaneously deploy a solution for the peer request problem and a solution for the service scheduling problem as well. That is, deploying either one of them separately does not yield the same attractive performance.

Thus, in order to achieve a more feasible overall solution, which cuts off the high message overhead as well as the performance interdependence between the above proposals, it is quite reasonable to simply consider the two following design concepts. First, the number of unchoked neighbours should be equal to the number of data upload slots. In this case, there will be no queued requests. Second, a peer selects the unchoking neighbours in the strict time order they unchoke it, i.e., the first neighbour to unchoke the peer is the first one to be selected, and so on. Thus, there is no decision for neighbour selection, since there is only one neighbour at a time.

## 4.2. Client's Heterogeneity

Among the works tackling the client's heterogeneity, the one presented by D'Acunto et al. [11] deserves special attention due to its wide-spectrum analysis. They propose an adaptive strategy for peers to decide, with respect to their current progress, whether relaxing the traditional peer-selection policy, based on direct reciprocity, and serve more random peers. Besides, they propose to have old peers giving preference to newly arrived peers. The implementation includes, for example, adjusting the number of data upload slots dynamically and taking into account the indirect-reciprocity principle. In general, the results show that low-capacity peers receive a satisfactory QoS, without affecting that of high-capacity peers. On the other hand, in case the bandwidth is scarce, best-uploaders still receive the highest-priority, as it happens under the traditional BitTorrent protocol. Even though this proposal may result in satisfactory QoS, its main drawback is the need for on-line computation. Besides, there is a dependence on indirect information due to the principle of indirect reciprocity.

## 4.3. Concurrent Data Upload Slots

The number of data upload slots a peer has determines the number of neighbours this peer may concurrently serve. Determining an optimal value to achieve high QoS is though an intricate task. Yang and De Veciana [14], for example, propose an analysis in transient and steady-state regimes. The analytical results and measurements chiefly suggest how various mechanisms might be designed to make a P2P system suitable for handling bursty and large data volume. *Branching process models* and *Markov chain models* are used in the analysis. Nevertheless, specific results for the exact optimal number of data upload slots for each peer are not derived.

Another example is the work of Biersack et al. [15]. They provide some more specific results, but no explicit conclusions for multimedia on-demand streaming scenarios are achieved. They conduct analytical studies based on different distributions models such as *linear chain architecture*, *tree distribution architecture*, and *parallel trees*. Among the most important findings, they notice that the number of data upload slots should be between three and five. In general, it may be observed that increasing the number of peers served at the same time has the side effect of decreasing the service rate to each peer.

From above, there has not been yet a definitive result concerning the optimal number of data upload slots for the various types of on-demand streaming systems. On the other hand, the interval between three and five slots may be probed to find out possible optimal values to each target application.

# 5. NOVEL PROPOSALS

The *Simple Interactive Streaming Protocol* (SISP) and the *Exclusive Interactive Streaming Protocol* (EISP) are introduced in this section. They are both explained by means of their corresponding peer-selection and piece-selection policies.

Algorithm 1 refers to the peer-selection policy of both proposals. It considers the point of view of a local peer that has just been told of its neighbours. Let $P_1$ and $P_2$ be two any swarm peers. Peer $P_1$ is *interested* in peer $P_2$ when it (peer $P_2$) has pieces that peer $P_1$ does not [16, 11, 6]. Peer $P_1$ is *not interested* in peer $P_2$ when it (peer $P_2$) has just a subset of the pieces peer $P_1$ does [16, 11, 6]. Moreover, *interested* means interested in the local peer, and *choked* means choked by the local peer. Comparing to the original BitTorrent, the difference mainly refers to the existence of the parameters $x_1$, $x_2$, $y$, $k$, and $\delta$. In general, the design concepts are: (i) the number of regular data upload slots is at least equal to the number of optimistic data upload slots; (ii) the direct-reciprocity principle is considered; and (iii) there is no on-line computation.

Algorithms 2 and 3 present the piece-selection policies of the SISP and EISP proposals, respectively. For objectivity, in Algorithm 3, only what differs from Algorithm 2 is presented. The descriptions consider that the local peer is up to request the next data piece. In general, their design concepts are: (i) use of a sliding window $W$ for high-priority pieces; (ii) use of a mixed piece selection, i.e., greedy for high-priority pieces, and rarest-first for low-priority pieces; (iii) there are no on-line computations; and (iv) use of a local client's buffer to avoid requesting the same data more than once.

---

**Algorithm 1: Peer-Selection Policy**

Begin
Initialization and definitions
    <u>Step 1</u> The neighbours of a local peer are choked and categorized in either *interested* or *not interested* remote peers.
    <u>Step 2</u> The local peer's upload capacity is divided into $x = (x_1 + x_2)$ upload slots: $x_1$ slots are *regular* slots, while $x_2$ slots are *optimistic* slots.
Repeat
    <u>Step 3</u> Every $\delta$ seconds (regular unchoke interval)
        <u>Step 3.1</u> If the local peer is in *leecher state*: all interested remote peers are ordered with respect to their upload rate to the local peer, and the $x_1$ fastest peers are unchoked. That is, the $x_1$ regular unchoke slots are assigned, one for each of the $x_1$ fastest peers.
        <u>Step 3.2</u> Else if the local peer is in *seed state*: all interested remote peers are ordered according to the upload rate the local peer has to them, and the $x_1$ fastest peers are unchoked.
    <u>Step 4</u> Every $k.\delta$ seconds (*optimistic unchoke interval*)
        Exactly $x_2$ additional interested remote peers are unchoked at random. That is, the optimistic unchoke slots are assigned.
Until the content download is complete.
End.

---

**Algorithm 2: Piece-Selection Policy of the SISP proposal**

Begin

Initialization and definitions

 <u>Step 1</u> The file data pieces are numbered and referenced this way: 1, 2, …, $t$. Each piece belongs to one category at a time: *high-priority* or *low-priority*. High-priority pieces are close in time to being played, while low-priority pieces are played in the future. A greedy policy is used to retrieve high-priority pieces, while a rarest-first policy is used for low-priority pieces.

 <u>Step 2</u> Let $d$ be the piece of the current playback point, and let $W$ be the sliding window. The window $W$ has length of $w$ data pieces. The first piece of $W$ is $d$, and its last piece is $(d + w)$. Pieces inside $W$ are high-priority, while pieces outside $W$ are low-priority.

 <u>Step 3</u> The window $W$ is updated in two cases. First, when pieces are normally played. Let $j_{play}$ be the number of pieces played. The first piece of $W$ is updated to $(d + j_{play})$, and its last piece to $(d + w + j_{play})$. Second, when there is a *Jump Forwards* (JF) or a *Jump Backwards* (JB). Let $j$ be the number of pieces skipped due to a JF or a JB. In case of a JF, the first piece of $W$ is updated to $(d + j)$, and its last piece to $(d + w + j)$; in case of a JB, the first piece of $W$ is updated to $(d − j)$, and its last piece to $(d + w − j)$.

Repeat

 <u>Step 4</u> Counting from the first piece of the sliding window $W$, if $v$ consecutive pieces have already being retrieved but have not been played yet, then request the rarest piece among all pieces falling outside $W$; otherwise, request the next piece falling inside $W$ until $v$ consecutive pieces have been retrieved but have not been played yet.

Until the content download is complete.

End.

---

**Algorithm 3: Piece-Selection Policy of the EISP proposal**

Repeat

 <u>Step 4</u> Counting from the first piece of the sliding window $W$, if $v$ consecutive pieces have already being retrieved, but have not been played yet, then request the rarest piece among all data pieces falling inside $W$. Otherwise, request the next piece falling inside $W$ until $v$ consecutive pieces have been retrieved but have not been played yet.

Until the file download is complete.

End.

---

In summary, it may be noticed from the previous algorithms that the EISP proposal basically differs from the SISP proposal in the following aspect: in EISP, the piece selection solely occurs inside the sliding window, while in SISP, this selection may occur inside and outside the sliding window. That is, while the EISP mostly prioritizes continuous playback, the SISP seeks a trade-off between continuous playback and data diversity.

## 6. PERFORMANCE EVALUATION

For the sake of an easier understanding and overall appraisal, this section is divided into three subsections. Subsection 6.1 explains the simulation scenarios explored. Subsection 6.2 chiefly focuses on the modelling as well as on the performance metrics used to evaluate the proposals. Lastly, Subsection 6.3 presents the results and corresponding analysis.

## 6.1. Scenario Characterization

To have a fair and realistic analysis, it is mandatory to consider distinct categories of replication scenarios. The rationale is that the scenarios may diverge from each other in a significant amount due to inherent specificities related to the content being replicated. For example, video content is much longer than song content and its playback is supposed to suffer more interactive actions. Being aware of this, Table 1 lists the main parameters of replication scenarios.

Table 1. Scenario characterization.

| Parameter | Notation | Definition |
|---|---|---|
| File type | $f_{type}$ | Category the file belongs to: Music Files, TV Series, Movies, and All Media. The AM category refers to all types of content. |
| File size | $f_{size}$ | Size of the file to be replicated. Measured in bytes. |
| Swarm size | $s_{size}$ | Number of peers involved in the file replication in steady state. |
| Number of seeds | $m$ | Number of peers that have all file pieces in steady state. |
| Upload rate | $r_{up}$ | Peer's rate to upload content. Measured in bytes per second. |
| Download rate | $r_{down}$ | Peer's rate to download content. Measured in bytes per second. |
| Piece size | $p_{size}$ | Logical data unit considered in the file replication process. Measured in bytes. |
| Block size | $b_{size}$ | Network data unit considered in the file replication process. Measured in bytes. |

Another important point refers to the client's interactive behaviour during the file playback. The scenarios admit the client may execute the following interactive actions: *Play*, *Stop*, *Pause*, *Jump Backwards* (JB) and *Jump Forwards* (JF). *Play* indicates the client plays the file at the normal playback rate. *Stop* means the playback is terminated. *Pause* is the state in which the playback is frozen. JF indicates the client moves his playback point forwards from his current playback point. Lastly, JB indicates the client moves his playback point backwards from his current playback point.

## 6.2. Simulation Setup and Performance Metrics

The proposals are devised on top of the Tangram-II modelling environment [23], which is an event-driven object-oriented simulation tool. Concerning this modelling, five points deserve to be outlined. First, the system is analyzed in steady state, i.e., although peers might join and leave the system, the number of peers remains constant. This means that the initial phase, which encompasses contacting the tracker, establishing TCP connections, and determining the neighbourhood, is not modelled. Second, the piece-selection policy of the original BitTorrent in fact consists of a set of other specific policies which are not within the scope of this work. These more specific policies map to three distinct stages, namely *initiating the download* (i.e., the random first policy), *normal operation* (i.e., the rarest first policy), and *pulling down the last* *remaining pieces* (i.e., the end game mode policy). The focus of this paper lies on the normal operation stage and hence it is the only part encompassed by the modelling.

Third, the interactive actions by the client are associated to the occurrence of an event $E$ of rate $\lambda$, as explained next. When event $E$ occurs, *Play*, *Pause*, JB, and JF are respectively triggered with probabilities $p_0$, $p_1$, $p_2$, and $p_3$. The *Stop* action is automatically triggered when the client's session terminates. The client's session refers to the time duration he stays in the system to receive the data he wants to, and it terminates when he receives the same amount of data corresponding to the total file size (hence, there is always only one seed during the simulations). Table 2 completes the description of the client's interactive behaviour. Fourth, the simulation results have 95% confidence intervals that are within 5% of the reported values, and the simulation time for each run is set to 1.0e+06 seconds.

Table 2. Interactivity parameters.

| Parameter | Notation | Definition |
|---|---|---|
| *Interactivity rate* | $\lambda$ | Rate at which event $E$ occurs. It determines the client's interactivity profile: High-Interactive, Medium-Interactive, and Low-interactive. Measured in number of events $E$ per client's session. It has a Poisson distribution. |
| *Play length* | $l_{play}$ | Size of the file segment that is played after *Play* is triggered by the client. Measured as a percentage of the file size. |
| *Jump length* | $l_{jump}$ | Size of the file segment that is skipped after JF (or JB) is triggered by the client. Measured as a percentage of the file size. |
| *Pause length* | $l_{pause}$ | Time interval the client remains at the same playback point after *Paus*e is triggered by him. Measured in seconds. |

As for the performance metrics, they are all explained in Table 3. Measuring each of them separately provides important evidences of the QoS level experienced by a system client. Nevertheless, it is the joint analysis of them that becomes more invaluable to obtain reliable conclusions for the efficient design of BitTorrent-like protocols.

Table 3. Performance metrics.

| Metric | Notation | Definition |
|---|---|---|
| *Number of Served Clients* | *CS* | Number of clients served in the simulation. This metric is mainly used to see whether an expressive number of clients have been served during the whole simulation, and if so the overall results do apply for a steady-state regime. |
| *Interruption Coefficient* | *IR* | Average number of missing data pieces during the client's file playback relativized to the total file size in number of data pieces. Numerically, it lies within the interval [0, 1]. The closer it is to 0, the more efficient the proposal is. |
| *Relativized Service Time* | *RST* | Average time needed by a client to receive the data solicited relativized to the total file size in number of data pieces. Mainly used to provide fair relative comparisons in terms of service time in replication scenarios of different file sizes. |
| *Retrieval Coefficient* | *RC* | Ratio between the average data-retrieving time through exclusive data channels (i.e., suffering no interruptions) and the average data-retrieving time through the swarm (i.e., likely suffering interruptions). It lies within the interval [0, 1]. The closer it is to 1, the more efficient the proposal is. |

## 6.3. Results and Analysis

This section is divided into three distinct subsections. Subsection 6.3.1 carries out a competitive performance evaluation between the novel proposals. Subsection 6.3.2 considers the best proposal and seeks to optimize it by determining ideal numerical values for the parameters in its algorithmic description. Subsection 6.3.3 examines scalability and the impact of interactivity.

The numerical values in Tables 4, 5, 6, and 7, which are used in the simulations, are all gleaned from recent and important literature works [3, 24–29], assuring an up-to-date analysis in near-real replication scenarios. For example, Souza e Silva et al. [26] found that about 73% of the swarms are formed by less than 10 peers, and 58% of them have less than five peers; Hoßfeld et al. [25] observed a relationship between the content type and the swarm size: for TV series, the average size is 15.53 peers per swarm and, for Movies and Music files, the averages are 25.46 and 9.76 peers, respectively, per swarm; Wang et al. [24] observed that only 30% of the non-video contents are larger than 100 MB, and over 50% of non-video contents are less than 20 MB, whereas those small contents are very few in the existing video file swarms; and the ratio between the maximum upload capacity and the real achieved average download rate may be significant, e.g., a ratio of six, or may even present high variability [27, 25]. To end, it is still worth mentioning that upper bound values are herein preferred to value intervals to generate a larger number of events, thereby enabling an extensive analysis and hence more reliable results.

Table 4. Replication scenarios.

| Parameters | Replication scenarios based on $f_{type}$ | | | |
| | All Media | Music Files | TV Series | Movies |
|---|---|---|---|---|
| $f_{size}$ | 20.0 MB | 10.0 MB | 100.0 MB | 200.0 MB |
| $s_{size}$ | 7 peers | 10 peers | 15 peers | 25 peers |

Table 5. Parameters with fixed values.

| Parameters | Numerical values |
|---|---|
| $p_{size}$ | 256 kB |
| $b_{size}$ | 16 kB |
| $r_{up}$ | 20 kB/s |
| $r_{down}$ | 20 kB/s |
| $m$ | 1 seed |

Table 6. Settings up.

| Parameters | Setting Up I (SU-I) | Setting Up II (SU-II) | Setting Up III (SU-III) |
|---|---|---|---|
| $x_1$ | 3 | 4 | 2 |
| $x_2$ | 1 | 1 | 2 |
| $k$ | 3 s | 3 s | 4 |
| $\delta$ | 10 s | 10 s | 10 s |
| $w$ | $l_{play}$ | $l_{play}$ | $l_{play}$ |
| $v$ | 50% of $l_{play}$ | 50% of $l_{play}$ | 50% of $l_{play}$ |

Table 7. Interactivity profiles.

| Parameters | Low Interactivity (LI) | Medium Interactivity (MI) | High Interactivity (HI) |
|---|---|---|---|
| $\lambda$ | 0.005/s | 0.014/s | 0.025/s |
| $l_{play}$ $l_{jump}$ | 14.5% of $f_{size}$ | 3.5% of $f_{size}$ | 1.5% of $f_{size}$ |
| $l_{pause}$ | (14.5% of $f_{size}$)/$r_{down}$ s | (3.5% of $f_{size}$)/$r_{down}$ s | (1.5% of $f_{size}$)/$r_{down}$ s |
| $p_0; p_1; p_2; p_3$ | 0.89;0.01;0.05;0.05 | 0.71;0.05;0.12;0.12 | 0.55;0.15;0.15;0.15 |

## 6.3.1. Overall Competitiveness

The two novel proposals are compared in this subsection. To this end, the four general scenarios in Table 4, the fixed values in Table 5, the Setting Up I in Table 6, and the client's interactivity profile of medium interactivity in Table 7 are considered.

Figures 1 and 2 plot the average values obtained for the metrics RC, IC, RST, and CS in distinct replication scenarios. On one hand, it may be observed that, under the All Media and Music Files scenarios, the performances of both proposals are quite similar, independently of the metric being observed. On the other hand, the EISP proposal clearly outperforms the SISP proposal in the TV Series and Movies scenarios. Considering the numerical values obtained in the case of the SISP proposal as reference, see that: the RC metric is 15.62% (for TV Series) and 11.64% (for Movies) greater in the case of the EISP proposal; the IC metric is 44.57% (for TV Series) and 32.62% (for Movies) smaller in the case of the EISP proposal; the RST metric is reduced at 12.90% (for TV Series) and 10.18% (for Movies) in the case of the EISP proposal. Lastly, the value of the CS metric does not change within a same scenario because the swarm size is set to be the same in steady-state regime for both proposals.

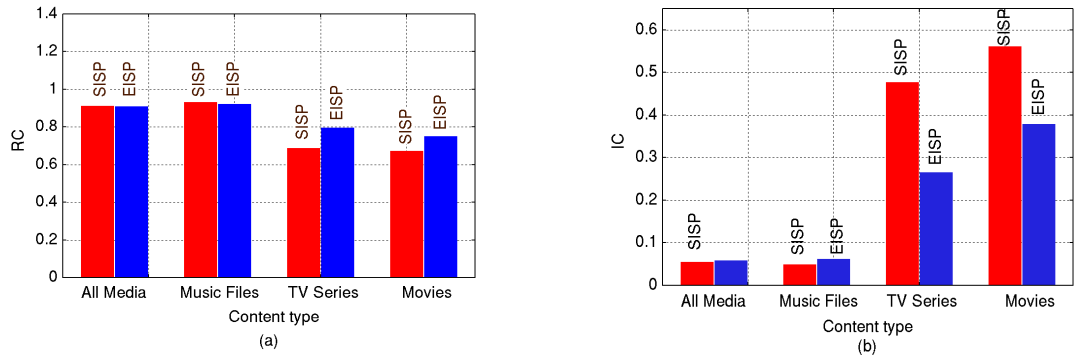

Figure 1. (a) RC in distinct scenarios; (b) IC in distinct scenarios.

Finally, note that the TV Series and Movies scenarios possess more competing peers and a larger file size to be replicated than the All Media and Music Files scenarios. Hence, the system complexity, in terms of message control and data structures manipulation, is higher in TV Series and Movies scenarios. So, for higher-complexity scenarios, prioritizing playback continuity is more efficient than prioritizing a trade-off between playback continuity and data diversity, since the EISP proposal outperforms the SISP proposal. Still, for lower-complexity interactive scenarios, either prioritizing data diversity or a trade-off between playback continuity and data diversity results the same, since the EISP and SISP proposals have very similar performances.

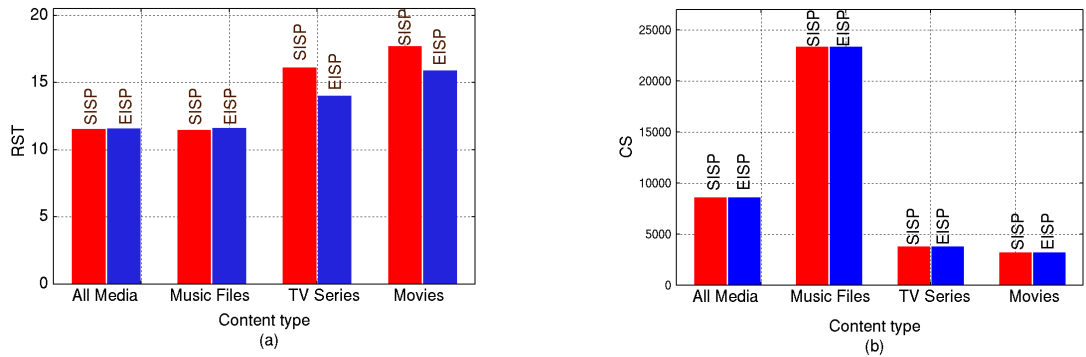

Figure 2. (a) RST in distinct scenarios; (b) CS in distinct scenarios.

**6.3.2. Optimization**

In this subsection, the focus lies on the EISP proposal, since it performs best in the last subsection. The goal is to see whether there may still be some performance optimization due to the use of other numerical values for the parameters in its algorithmic description. To this end, the Movies scenario in Table 4, the values in Table 5, the three Settings Up in Table 6, and the client's interactivity profile of Medium Interactivity in Table 7 are considered.

Considering the above, Table 8 brings the average values of RC, IC, RST and CS related to the EISP proposal. Comparing the average values obtained for a same metric, the variance of RC, IC, RST, and CS are $0.6999e{-}07$, $0.5470e{-}07$, $0.0152$, and $184.3333$, respectively. From this, it may be seen that there is no noticeable change among the distinct settings up. Thus, neither increasing the number of data upload slots (in SU-II) nor equally sharing bandwidth capacity (regular and optimistic slots) plus increasing the optimistic unchoke interval (in SU-III) may optimize the EISP proposal. Even though this observation is obtained for interactive on-demand streaming scenarios, it wholly agrees with the more-general one achieved in the work of Biersack et al. [15] (see Subsection 4.3.).

Table 8. Results for optimization analysis.

| Settings Up | Performance metrics | | | |
|---|---|---|---|---|
| | RC | IC | RST | CS |
| SU-I | 0.7483 | 0.3771 | 15.8595 | 3136 clients |
| SU-II | 0.7484 | 0.3770 | 15.8672 | 3135 clients |
| SU-III | 0.7479 | 0.3730 | 16.0770 | 3112 clients |

Lastly, using over five data upload slots to optimize the EISP proposal would be an intriguing and more detailed analysis. The modelling should make download/upload rates function of the number of data upload slots each peer opens. Let $B$ be the maximum peer's bandwidth capacity, and $S$ the total number of slots this peer opens. Each slot would then have a bandwidth capacity computed by $B/S$. The expected final result would therefore be a set of pairs of values $(B_i, S_i)$, where $S_i$ would stand for the ideal number of slots for a given bandwidth capacity $B_i$. This analysis though lies out of the scope of this paper and is encouraged for future work.

**6.3.3. Scalability and Interactivity**

This subsection has two prime goals. The first is to determine how scalable the EISP proposal is, and the second is to evaluate the impact of the client's interactive behaviour on the final performance. By scalable, it is herein meant that the proposal is able to handle a growing number of competing peers without deteriorating the final QoS. Measuring the defined performance metrics is a practical and efficient way to accomplish both goals.

The Movies scenario in Table 4, the fixed values in Table 5, the Setting Up I in Table 6, and the client's interactivity profile of medium interactivity in Table 7 are considered to achieve the first goal. The number of clients in steady state is considered to be 25, 40, and 50 at a time. The upper limit, set to 50, does suffice, since it emulates a scenario that at least doubles the most common values observed for real replication scenarios of this nature. Finally, Table 9 shows the final simulation results (average values) for this analysis.

Table 9. Results for scalability analysis.

| Swarm sizes | Performance metrics | | | |
|---|---|---|---|---|
| | RC | IC | RST | CS |
| 25 peers | 0.7483 | 0.3771 | 15.8595 | 3136 clients |
| 40 peers | 0.7513 | 0.3756 | 15.7596 | 5040 clients |
| 50 peers | 0.7523 | 0.3755 | 15.7492 | 6310 clients |

Comparing the values obtained for a same metric in Table 9, the variance of RC, IC, and RST are $4.4894e-06$, $8.0333e-07$, and $3.7076e-03$, respectively. From this, it is clear that there is no noticeable change for different swarm sizes and hence the final performance is satisfactory in all of them. Although it is true that the total number of clients served (CS) changes, presenting a variance of $255.2065+04$, this is already expected since the number of clients in steady state is increased at 100%, i.e., from 25 to 50 peers. Thus, the EISP is indeed a very scalable solution.

Interestingly, these results also show that the population growth, from 25 to 50 peers, does not increase the eventual total system capacity indirectly reflected by the final perceived QoS. Even though each new peer joining the system brings additional resources to the system, there seem to be intrinsic limitations of the P2P paradigm to cope with the population growth above a certain threshold. This observation leads to delimit two distinct performance regions: one where the system capacity scales in proportion to the population growth, and another one where there is stabilization or maybe degradation [30–32]. This discussion is left for future work.

Figures 3 and 4 refer to the second goal. They depict the average values obtained for the metrics RC, IC, RST, and CS for the three interactivity profiles in Table 7. It may be seen that the more interactive the client is, the better the performance becomes. The rationale is that a more interactive client tends to access the same file parts more often than a less interactive client does, while a less interactive client tends to access more distinct file parts than a more interactive client does. Since all retrieved data is permanently stored in the client's local buffer, interruptions due to missing pieces are hence more likely to occur for less interactive clients than for more interactive clients. This surely highlights the importance of the local buffer.

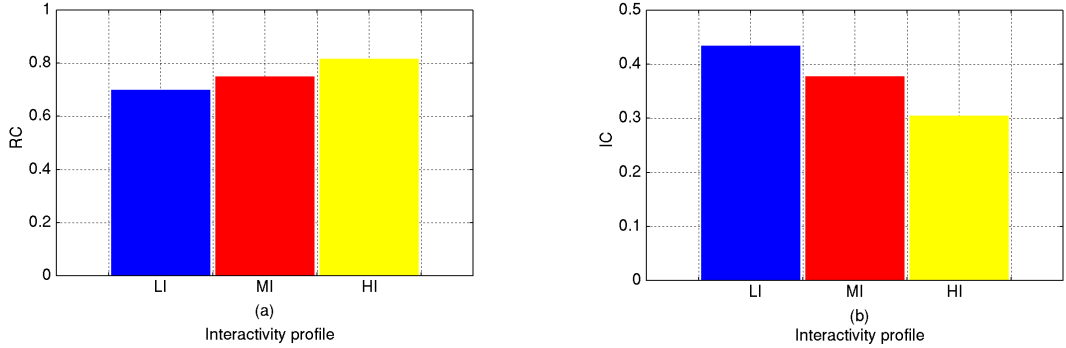

Figure 3. (a) RC for distinct interactivity profiles; (b) IC for distinct interactivity profiles.

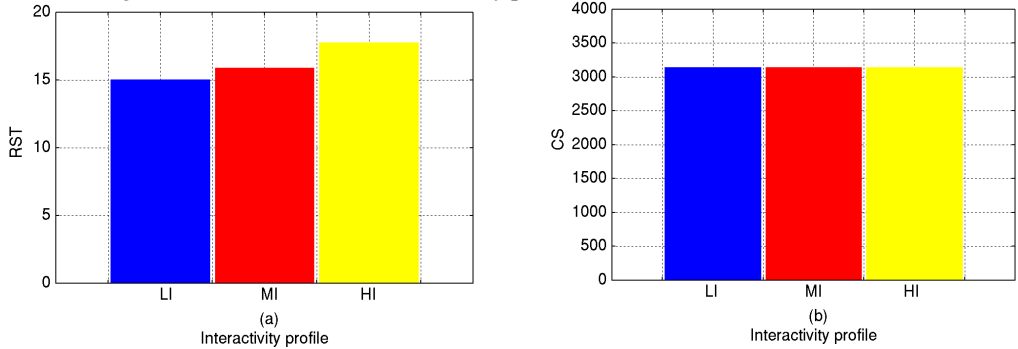

Figure 4. (a) RST for distinct interactivity profiles; (b) CS for distinct interactivity profiles.

Still, in Figures 3 and 4, when moving from LI towards HI, see that: the value of RC increases 16.81%; the value of IC decreases 29.78%; the value of RST increases 18.13%, confirming that a more interactive client stays in the system for a longer time than a less interactive client to retrieve the same quantity of data; lastly, the value of CS in all cases reaches up to 3.1360e+03, thereby indicating that the total number of served clients is significant and so is the number of events in the simulations.

# 7. CONCLUSION AND FUTURE WORK

This work presented two BitTorrent-like protocols for interactive multimedia streaming: the *Simple Interactive Streaming Protocol* (SISP) and the *Exclusive Interactive Streaming Protocol* (EISP). Their chief design concepts were based on prioritizing either playback continuity or a trade-off between playback continuity and data diversity. Related work and open issues in the realm of BitTorrent-like systems were reviewed in order to assure a thorough and up-to-date approach. The performance evaluation was carried out through simulations in a variety of file replication scenarios, assessing distinct performance metrics.

Some of the main conclusions achieved were: despite the fact that both proposals showed to be pretty efficient and scalable, prioritizing playback continuity was the best design concept, since it enabled the most optimized final results; the number of data upload slots and the unchoking time intervals of the original BitTorrent protocol were enough in order to have the novel proposals with optimized performances; lastly, the more interactive the scenario was, the more efficient the novel proposals became, highlighting the great importance of deploying a client's local buffer to avoid requesting the same data more than once.

Finally, future work may include the following avenues: to propose novel analytical and simulation models to more accurately understand how P2P system capacity scales in proportion to the population growth, including systems under flash crowds [30–32]; envisioning more advanced networks, to consider other replication scenarios with much larger file sizes and download/upload rates than those examined herein [24–29]; to compare the paradigms of P2P systems and Information Centric Networks (ICN) [33], respectively, to find out which one is potentially more resilient to the unprecedented increase of global traffic in today's Internet; and to develop novel P2P algorithms devoted to live streaming and mobile networks [18, 34].

## ACKNOWLEDGEMENTS

We are in debt to Doctor Guilherme Dutra Gonzaga Jaime from the Nuclear Engineering Institute, Brazil, for patiently answering our questions about the simulation tool Tangram.

## REFERENCES

[1]     Cohen, B. (2003) "Incentives build robustness in BitTorrent", *First Workshop on Economics of Peer-to-Peer Systems*, Berkeley, EUA.

[2]     D'Acunto, L., Chiluka, N., Vinkó, T. & Sips, H. (2013) "BitTorrent-like P2P approaches for VoD: A comparative study", *Computer Networks*, Vol. 57, No. 5, pp 1253 – 1276.

[3]     Hoffmann, L. J., Rodrigues, C. K. S. & Leão, R. M. M. (2011) "BitTorrent-like protocols for interactive access to VoD systems", *European Journal of Scientific Research*, Vol. 58, No. 4, pp 550 – 569.

[4]     Rodrigues, C. K. S. & Leão, R. M. M (2007) "Bandwidth usage distribution of multimedia servers using Patching", *Computer Networks*, Vol. 51, No. 3, pp 569 – 587.

[5]     Hua, Kai-Lung, Chiu, Ge-Ming, Pao, Hsing-Kuo & Cheng, Yi-Chi (2013) "An efficient scheduling algorithm for scalable video streaming over P2P networks", *Computer Networks*, Vol. 57, No. 14, pp 2856 – 2868.

[6]     Rodrigues, C. K. S. (2014) "Analyzing peer selection policies for BitTorrent multimedia on-demand systems in Internet", *International Journal of Computer Networks & Communications (IJCNC)*, Vol. 6, No. 1, pp 203 – 221.

[7]     Ramzan, N., Park, H. & Izquierdo E. (2012) "Video streaming over P2P networks: Challenges and opportunities", *Signal processing: Image Communication*, Vol. 27, pp 401 – 411.

[8]     Romero, P., Amoza, F. R. & Rodríguez-Bocca, P. (2013) "Optimum piece selection strategies for a peer-to-peer video streaming platform", *Computer and Operations Research*, Vol. 51, No. 5, pp 1289 – 1299.

[9]     Streit, A. G. & Rodrigues, C. K. S. (2013) "On the design of protocols for efficient multimedia streaming over Internet", *ESPE – Ciencia y Tecnología*, Vol. 4, No. 1, pp 25 – 39.

[10]    Yang, Y., Chow, A., Golubchik, L. & Bragg, D. (2010) "Improving QoS in BitTorrent-like VoD systems", *Proceedings of the IEEE INFOCOM*, San Diego, CA, USA.

[11]    D'Acunto, L., Andrade, J. & Sips, H. (2010) "Peer selection strategies for improved QoS in heterogeneous BitTorrent-like VoD systems", *IEEE International Symposium on Multimedia*, Taichung, Taiwan.

[12]    Mol, J., Pouwelse, J., Meulpolder, M., Epema, D. & Sips, H. (2008) "Give-to-Get: Free-riding-resilient video-on-demand in P2P systems", *SPIE MMCN*, San Jose, California, USA.

[13]    Liao, Wei-Cherng, Papadopoulos, F. & Psounis, K. (2007) "Performance analysis of BitTorrent-like systems with heterogeneous users", Performance Evaluation, Vol. 64, No. 9-12, pp 876 – 891.

[14]    Yang, X. & De Veciana, G. (2004) "Service capacity of peer-to-peer network", *23rd Annual Joint Conference of IEEE Computer and Communications Societies (INFOCOM 2004)*, Vol. 4, pp 2242 – 2252.

[15]    Biersack, E., Rodriguez, P. & Felber, P. (2004) "Performance analysis of peer-to-peer networks for file distribution", *Lectures Notes in Computer Science (LNCS)*, Vol. 3266, pp 1 – 10.

[16]    Legout, A., Urvoy-Keller, G. & Michiardi, P. (2006) "Rarest first and choke algorithms are enough", *6th ACM SIGCOM Conference on Internet Measurement*, Rio de Janeiro, RJ, Brazil.

[17]    Vlavianos, A., Lliofotou, M. & Faloutsos, M. (2006) "BiToS: Enhancing BitTorrent for supporting streaming applications", *25th IEEE International Conference on Computer Communications (INFOCOM 2006)*, Barcelona, Spain, pp 1 – 6.

[18]    Zhang, X. & Hassanein, H. (2012) "A survey of peer-to-peer live video streaming schemes – An algorithmic perspective", *Computer Networks*, Vol. 56, No. 15, pp 3548 – 3579.

[19]    Rocha, M. V. M. & Rodrigues, C. K. S.  (2013) "On Client's interactive behaviour to design peer selection policies for BitTorrent-like protocols", *International Journal of Computer Networks & Communications (IJCNC)*, Vol. 5, No. 5, pp 141 – 159.

[20]    Borghol, Y., Ardon, S., Carlsson, N. & Mahanti, A. (2010) "Toward efficient on-demand streaming with BitTorrent", *IFIP Networking*, Chennai, India.

[21]    Shah, P. & Pâris, J.-F. (2007) "Peer-to-Peer Multimedia Streaming Using BitTorrent", *IEEE International Performance, Computing, and Communications Conference (IPCCC)*, New Orleans, Louisiana, USA, pp 340 – 347.

[22]    Carlsson, N. & Eager, D.   (2007) "Peer-assisted on-demand streaming of stored media using BitTorrent-like protocols", *IFIP Networking*, Atlanta, GA, USA.

[23]    De Souza e Silva, E., Figueiredo, D. & Leão, R. (2009) "The Tangram-II Integrated Modelling Environment for Computer Systems and Networks", *ACM SIGMETRICS Performance Evaluation Review*, Vol. 36, No. 4, pp 45 – 65.

[24]    Wang, H., Liu, J. & Xu, K. (2012) "Understand traffic locality of peer-to-peer video file swarming", *Computer Communications*, Vol. 35, No. 15, pp 1930 – 1937.

[25]    Hoßfeld, T., Lehrieder, F.,  Hock, D., Oechsner, S., Despotovic, Z., Kellerer, W. &  Michel, M. (2011) "Characterization of BitTorrent swarms and their distribution in the Internet", *Computer Networks*, Vol. 55, No. 5, pp 1197 – 1215.

[26]    De Souza e Silva, E., Leão, R., Menasché, D. & Rocha, A. (2013) "On the interplay between content popularity and performance in P2P systems", *Lecture Notes in Computer Science (LNCS)*, Vol. 8054, pp 3 – 21.

[27]    Murai, F., Rocha, A., Figueiredo, D. & Souza e Silva, E. (2011) "Heterogeneous download times in a homogeneous BitTorrent swarm", *Computer Networks*, Vol. 56, No. 7, pp 1983 – 2000.

[28]    Ye, L., Zhang, H., Li, F. & Su, M. (2010) "A measurement study on BitTorrent system", *International Journal of Communications, Network and System Sciences*, Vol. 3, pp 916 – 924.

[29]    Varvello, M., Steiner, M. & Laevens, K. (2012) "Understanding BitTorrent: a reality check from the ISP's perspective", *Computer Networks*, Vol.56, No. 40, pp 1054 – 1065.

[30]    Zhu, J. & Hajek, B. (2012) "Stability of a peer-to-peer communication system", *Information Theory, IEEE Transactions*, Vol. 58, No. 7, pp 4693 – 4713.

[31]    Menasché, D., Rocha, A., De Souza e Silva, E., Towsley, D. & Leão, R. (2012) "Implications of peer selection strategies by publishers on the performance of P2P swarming systems", *ACM SIGMETRICS Performance Evaluation Review*, Vol. 39, No. 3, pp 55 – 57.

[32]    De Souza e Silva, E., Leão, R., Menasche, D. & Towsley, D. (2014) "Scalability Issues in P2P Systems", *arXiv preprint arXiv:1405.6228*.

[33]    Domingues, G., De Souza e Silva, E., Leão, R. & Menasché, D. (2013) "Enabling Information Centric Networks through opportunistic search, routing and caching", *31st Brazilian Symposium of Computer networks and Distributed Systems (SBRC)*, Brasília, DF, Brazil.

[34]    Gulati, M. & Kumar, K.  (2014)  "Performance comparison of mobile Ad Hoc network routing protocols", *International Journal of Computer Networks & Communications (IJCNC)*, Vol. 6, No. 2, pp 127 – 142.

**Author**

Carlo Kleber da S. Rodrigues received the B.Sc. degree in Electrical Engineering from the Federal University of Paraiba in 1993, the M.Sc. degree in Systems and Computation from the Military Institute of Engineering (IME) in 2000, and the D.Sc. degree in System Engineering and Computation from the Federal University of Rio de Janeiro (UFRJ) in 2006. Currently he is Military Assessor of the Brazilian Army in Ecuador, Professor at the Armed Forces University (ESPE) in Ecuador, and Professor at the University Center (UniCEUB) in Brazil. His research interests include the areas of computer networks, performance evaluation, and multimedia streaming.

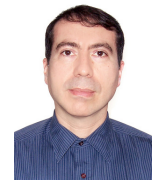